\documentclass[conference]{IEEEtran}

\usepackage{cite}
\usepackage{color}
\usepackage{verbatim}
\usepackage[cmex10]{amsmath}
\usepackage{url}

\usepackage{cite}
\usepackage{ifthen}
\usepackage{epsfig, times, subfigure}
\usepackage{amssymb,amsmath,amsfonts}
\usepackage{graphicx}
\usepackage{array}

\newtheorem{definition}{Definition}

\begin{document}
\title{Anatomy of the Internet Peering Disputes}

\author{\IEEEauthorblockN{Siddharth Bafna}
\IEEEauthorblockA{The LNMIIT\\
Jaipur, India\\
Email: bafna.siddharth@gmail.com}
\and
\IEEEauthorblockN{Avichal Pandey}
\IEEEauthorblockA{The LNMIIT\\
Jaipur, India\\
Email: pandeyavichal7@gmail.com}
\and
\IEEEauthorblockN{Kshitiz Verma}
\IEEEauthorblockA{Universidad Carlos III de Madrid\\
The LNMIIT, 
Jaipur, India\\
Email: vermasharp@gmail.com}
}

\maketitle

\begin{abstract}
Internet peering disputes have had an impact on the Internet AS-graph. As a result, the customers of the ASes often suffer because they cannot reach to the all of the Internet. There is a lack of study of the disputes that have taken place so far, even though each dispute is individually well understood. In this paper, we collect data on 26 disputes from from various resources, categorize them in a systemic manner to understand them from geographical and temporal point of views. There are some ASes that are more involved in disputes than others. In the end, we conclude we need to collect more data as it would be more interesting to have data on the Internet peering disputes around the world. 
\end{abstract}
\IEEEpeerreviewmaketitle

\section{Introduction}
The Internet can arguably be called as the most spectacular technological artifact of our times that has emerged from a distributed, uncoordinated, spontaneous interaction of many \cite{fabrikant2003network} and the rapid emergence of the Internet has even amazed the people responsible for its creation. It has shrunk the world beyond the limitations posed by politics, culture, and physical boundaries of nations. Its importance is evident from the fact that the United Nations have considered making access to the Internet as human right. The unit of a network is an Autonomous System (AS). Today's Internet comprises of around 65,516 (as of March 30, 2014) Autonomous Systems (ASes) \cite{asn} administering thousands of network computing devices. All the ASes are connected to each other using the Border Gateway Protocol version 4 (BGP) to provide connectivity to the end users.

Business relationships among Autonomous Systems is essential for the connectivity provided by the Internet, which is mainly of two kinds. 1) \emph{Transit}: Relationship wherein one AS acts as a customer and takes services, of what is called its provider, to get the connectivity to the rest of the Internet. To ensure connectivity to the Internet, establishing transit relationships is necessary and the customer must pay monthly bills to the provider. 2) \emph{Peering}: When two ASes see a mutual benefit to save the transit costs, i.e., instead of paying to the transit provider, they establish a connection among themselves to exchange each others traffic. Note that it can be established only among two ASes such that none can be the provider of the other. If one could be provider, it will not engage in payment free peering.

Peering can further be classified into two categories, public and private Peering. IXPs (Internet Exchange Points) facilitate connection between two ASes by providing them the necessary infrastructure. When two ASes peer at an IXP, it can be termed as public peering. A peering is private if two ASes deploy their own infrastructure to connect to each other. It is often cost effective to peer at IXP because all the necessary infrastructure is already there.

Since peering happens for mutual benefit, none of the ASes can take unfair advantage. For example, one AS may inject much more traffic on the peering link than the other. If is often measured as peering traffic ratio, which is defined as out:in and should not exceed 2:1. Any imbalance often results in what is known as peering dispute.

\begin{definition}[peering dispute]
\label{dispute}
We classify all the events as disputes in which a conflict between two peers leads to the termination or warning of termination of the peering link, from any one of them, even for a short period of time. 
\end{definition}

It may be noted that 500 to 1000 small de-peerings occur daily \cite{cs.columbia}, but we do not consider them as disputes. These de-peerings mainly occur due to technical incapabilities or negligence in maintaining peering links using BGP. So, we consider a de-peering event as a dispute only when there is a conflict of interest between the parties or there is a claim made for the breach of the peering agreement from at least one of the involved parties.

Peering disputes are not mare concern of the two involved ASes. It affects a lot bigger region depending upon the AS. An AS is in tier-1 if it has no transit provider. Hence, it does not pay to anyone for transit but rather forms a peering mesh with the other tier-1 ASes. Hence, as we will see later in the paper, depeering can fragment the Internet and leave the end users disconnected from the other part. It impacts BGP routing and its convergence. 

\subsection{Our Contribution}
In this paper, we aim to understand the Internet peering disputes that disturb the connectivity of the Internet and often result in end users sufferings. We collected disputes from various sources including blogs, news, mailing lists like NANOG, etc. We do a first detailed analysis of the nature of such disputes. Some of our contributions can be summarized as:
\begin{itemize}
\item We collected 26 Internet peering disputes that have occurred since the Internet was commercialized, including the ASes involved, reason and duration of disputes. 
\item We give a general classification for the disputes in which they can be categorized.
\item We give temporal distribution of disputes since its commercial inception until today and geographical distribution according to different Internet Regions.
\end{itemize} 
Such a study is crucial in understanding the evolution of the Internet and predict a dispute that may happen in future.

The rest of the paper is organized as follows: Section \ref{relwork} discusses the related work, Section \ref{dc} describes the methodology behind our data collection and explains the collected data, Section \ref{results} presents the inferences and Section \ref{conclusion} concludes the paper with potential future work to be done.

\begin{table*}[t]
\label{dt}
\centering
\caption{List of Peering Disputes Collected}
\begin{center}
    \begin{tabular}{ | l | l | l | p{8.5 cm} |}
    \hline
    {\textbf{S.No.}} & {\textbf{Conflicting Companies}} & {\textbf{Month/Year}} & {\textbf{Reason}} \\ \hline
        1. & Telecom Italia - Other ISPs & July'13 & Telecom Italia was reducing the 	number of neutral access points \\ \hline
	2. & Cogent - Verizon & June'13 & Verizon neglected upgrading the peering connection \\ \hline
	3. & FT Orange - Cogent + Google & Jan'13 & FT-Orange restricted bandwidth for online video service Youtube \\ \hline
	4. & Cogent - China Telecom & Mar'12 & Parties de-peered for unknown reasons \\ \hline 
	5. & Cogent - France Telecom & Aug'11 & France Telecom didn't allow Cogent to connect with its Customers \\ \hline
	6. & Cogent  - ESNet & June'11 & ESNet was below the Cogent's minimum traffic volume threshold \\ \hline
	7. & Level3 - Comcast & 2010 & Comcast started charging new fee to deliver Level3 traffic \\ \hline
	8. & Cogent - Hurricane Electric & Oct'09 & Both are IPv6 Tier 1 backbone, cogent de-peered HE \\ \hline
	9. & Chunghwa Telecom - TFN  & Apr'09 & Reason not known \\ \hline
	10. & Sprint - Cogent & Sept'08 & Traffic Exchange Criteria not met \\ \hline
	11. & Telia - Cogent  & Mar'08 & Imbalanced Traffic Ratios \\ \hline
	12. & Cogent - Limelight & Sept'07 & Cogent de-peered Limelight for unknown reasons \\ \hline
	13. & Cogent - Level3 & Oct-05 & Link Terminated due to imbalanced Traffic Ratio \\ \hline 
	14. & AOL - MSN & Sept'03 & Reasons unknown, but AOL users were not able to access MSN \\ \hline
	15. & Cogent - AOL & Dec'02 & Imbalanced Traffic Ratio \\ \hline
	16. & C\&W - PSINet & 2001 & C\&W dropped the peering agreement \\ \hline
	17. & BBN/Genuity/GTE - Exodus & Before 2001 & Battle over imbalanced traffic flows \\ \hline 
	18. & BBN/GTE - MCI/Worldcom & Around '99 & Nature of peering agreement was not clarified \\ \hline
	19. & UUNet – Whole Earth Networks Inc  & May'97 & UUNet demanded for paid peering \\ \hline
	20. & UUNet- Others & May'97 & UUNet notified its peers that they would terminate their peering \\ \hline 
	21. & AGIS - Others & Before '97 & AGIS announced its new peering policy at the NANOG meeting \\ \hline 
	22. & Digex Inc - AGIS & Oct'96 & Reasons not known \\ \hline
	23. & Sprint - Other ISPs &  Before '96 & Sprint refused to upgrade its connection at the CIX router \\ \hline
	24. & BBN - Other ISPs & Around '95 & BBN terminated its connection at CIX router \\ \hline 
	25. & BBN - ANS & Around '95 & BBN broke the agreement \\ \hline 
	26. & DANTE - EUNet & Oct'94 & DANTE asked EUnet to increase their connection rate \\ \hline 	
    \end{tabular}
\end{center}
\end{table*}

\section{Related work}
\label{relwork}
Internet peering relationships have always been the source of confusion among researchers. However, a very little about them is known \cite{ager2012anatomy}. Moreover peering involves skills beyond technology \cite{norton2002art}. Peering disputes have been talked about in literature before \cite{labovitz2001shining} \cite{ergas2000internet} and their importance acknowledged \cite{lodhi2012genesis} \cite{ma2008interconnecting}. As far as we know, there is not much in the literature on concretely studying the disputes themselves. Our work presents the first work in such a direction.

\section{Data Collection}
\label{dc}

In this section, we describe the methodology to gather peering dispute data. We describe the data that we have collected and later classify it according to  different disputes.

\subsection{Methodology}
\label{methodology}
In the absence of any literature, we started exploring news and mailing lists like NANOG. By the time we had finished, we had explored the Internet for peering disputes. We found maximum disputes in the region of North America, which was expected. We searched the mail archives of all the Internet Regional Registries. We also posted mails on discussion forums of all the Internet Regional Registries (NANOG, APNIC, RIPE NCC, AfriNIC and LACNIC) asking for more information on the peering and peering disputes we collected and also some new disputes. To widen our search for disputes in other parts of the world we used keywords in regional languages for our search. We translated the keywords that could potentially give us news on depeerings in local news. For example, we searched in Spanish and Portuguese in Latin America, Italian in Italy, French in France and so on.

\subsection{Data Collected}
In total we collected information about 26 peering disputes from all our sources. We were not surprised to collect only 26 disputes because most of the Internet Interconnection Agreements are very quite affair and are not documented for, they are mostly handshake agreements where parties mutually agree without any on record documentation. This argument is supported by the fact that 141,512 Internet Interconnection Agreements out of 142,210 Internet Agreements examined till March 2011 were Handshake Agreements \cite{woodcock2011survey}.


In TABLE I we have listed all the information on different peering disputes we collected, in the descending order on the basis of Month/Year. In these 26 collected disputes we have covered a wide spectrum of the Internet, from the early commercial phase to the current phase, the earliest peering dispute in our table is of Oct'94 and the most recent one is of July'13. Here, we would like to mention that we collected information on a more recent dispute anticipation \cite{cogent-battle} but, we have not included this in our list because this is just a prediction.

In the table, we have mentioned the reason for all the collected disputes barring a few, the reasons we mentioned are mostly based on the public statements made by the concerned companies. 

In some of the peering disputes collected, we could not find the exact month/year so we have mentioned the period based on some evidences. For example, dispute no. 23, 24 and 25 have their root at the CIX router (Commercial Internet eXchange, first IXP in USA). CIX was established in '91 and NSFNET reverted back to research network in '95 \cite{h2004hobbes}. Dispute no. 21 happened because AGIS announced its new peering policy at the NANOG meeting held at the University of Michigan \cite{nanog}, possibly NANOG 8 held in Oct'96 \cite{nanog8}. The period of dispute no. 17 and 18 was estimated based on the fact that exodus (involved in 17) filed for bankruptcy in 2001 \cite{exodus} and MCI/Worldcom (involved in 18) merged in '98 \cite{mci}. 

\begin{figure}
\centering
\includegraphics[width=3.0in]{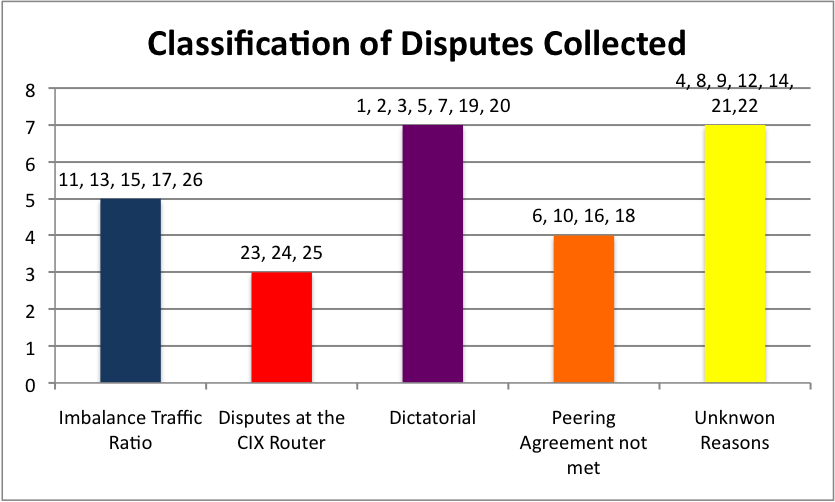}
 \caption{Classification of the Peering Disputes Collected}
\label{fig_class}
\end{figure}

\subsection{Classification of disputes}
Based on the similarities between the nature of disputes, we classifying them as mentioned below and as shown in Fig. \ref{fig_class}. It is possible that one dispute may belong to more than one classification but we put them in exactly one, giving the reason for our choice. 
\subsubsection{Imbalance Traffic Ratio}
Dispute no. 11, 13 and 15 are similar kind of disputes, the reason for these disputes is breach of imbalanced traffic ratio, i.e. one of the player accused the other of delivering more traffic in the peering link than the agreed limit without sharing the fair cost. Here, we have to consider that the exchange of traffic in the peering link is free of cost until an agreed limit. If this limit is exceeded then this simply means that one of the companies is taking an unfair advantage of this peering and this leads to the dispute \cite{brown2008peering} \cite{level3-cogent} \cite{cogent-aol}. Note that dispute no. 17 is also because of this same reason \cite{nanog} but there is a slight difference in the nature of traffic, in dispute no. 17 the nature of traffic is mostly web and peer-to-peer and in the other three the nature of traffic is mostly content of Internet giants like Google and Netflix which is requested by the end user. Another dispute based on a link over-use is number 26 between DANTE and EUNet. DANTE provided EUnet with a free 64 kbps access, but in practice much more capacity was used and due to this reason DANTE asked EUnet to increase their connection rate accordingly \cite{dante-eunet}.

\subsubsection{Disputes at the CIX Router}
There are three disputes in this category. CIX acted as the central point of connection for all the ISPs then. The battle between BBN and ANS was for the reason that BBN broke the peering agreement and this in turn lead to the connection of ANS at CIX Router. This peering dispute can also be classified in the Peering Agreement section but we kept it in this section because this is a dispute of a period when peering agreements were not so well defined, i.e. there was no exact format of a peering agreement and even publicizing of agreement details, which is considered as a violation in present time was not considered as a breach of the agreement due to the lack of the non disclosure agreements as mentioned in \cite{nanog}. The other battle in which BBN was involved was the termination of link at CIX. BBN was one of the first to do so and this trend continued and lead to the elimination of CIX as the router of last resort. The battle of Sprint with other ISPs lead to a situation of packet loss. Sprint neglected to upgrade its connection to the CIX Router and this made things painful for others because traffic exchange rate with Sprint was fairly low compared to the others \cite{nanog}.

\subsubsection{Dictatorial}
Considering the dispute no. 3, 5 and 7, these disputes are same in some way to the disputes discussed above, the reason for these disputes is also imbalance traffic ratio but the difference is that in all of these disputes the big player tried to use its customer base as an leverage directly or indirectly. Level3 and Comcast had their fight over the content of Netflix, Level3 was responsible for delivering the Netflix content, this increased the incoming traffic in the peering link for Comcast and due to this Comcast started charging a new fee on Level3 to deliver this content to its end users. The argument given by Comcast was that Level 3 was trying to send heavy traffic across its network without bearing its fair share of the cost, indirectly this was an attempt by Comcast to get a share of revenue which Level3 generated for acting as a Content Distribution Network (CDN) of the Netflix. This problem was however resolved on unknown mutually satisfactory terms \cite{level3-comcast}. Similar things happened in the two disputes between France Telecom (now Orange S.A.) and Cogent, one in Aug'11 \cite{cogent-orange} and the other in Jan'13 \cite{cogent-ft}. In the first dispute, Orange didn't allow Cogent to connect with its customers in France for free, citing the reason that Cogent was over loading the peering link by delivering third party content to the end users of France Telecom and in the second conflict Google was indirectly involved due to its video streaming product YouTube, Orange asked Cogent to pay for the additional traffic being generated by streaming video services. In all of these disputes the problem was due to third party content because Level3 and Cogent are both content delivery networks of Netflix and Google and the other parties involved in the disputes want a share of the revenue generated by Level3 and Cogent for delivering the heavy content traffic to their end users. Another conflict of this kind is no. 2 between Cogent and Verizon, the peering link between the two started to overflow because Verizon neglected to upgrade the peering connection. One possible reason for this as mentioned in \cite{cogent-verizon} is that most of the incoming traffic was coming from the CDN of Netflix picked up by Cogent and Verizon claimed that this was simply the violation of peering terms.


Dispute no. 1 between Telecom Italia and Other Italian ISPs was because Telecom Italia was reducing the number of neutral access points, the reason for doing this is not known to us but this de-peering lead to slow and poor quality connections over the Internet \cite{italia}. The other two disputes collected in this category involve UUNet\footnote{Now operates under Verizon Communications}, in the first one with Whole Earth Networks Inc, UUNet demanded for paid peering, Whole Earth first resisted this move citing abuse of market power by UUNet but ultimately agreed to paid peering \cite{cukier1998peering} and the second was because UUNet notified other peering providers that they would terminate their peering connection.\cite{nanog}  

\subsubsection{Peering Agreements not met}
By now we have established that a peering relationship is usually between equal players and for a peering to continue without any glitches, it is important for both the parties to follow the peering agreement and the key point of this agreement is the traffic ratio. Until now we explained the disputes where the reason was heavy traffic but there are disputes where de-peering happened because minimum traffic exchange criteria was not met. Sprint de-peered Cogent in Sept'08 \cite{sprint-cogent} and Cogent de-peered ESNet in June'11 \cite{cogent-esnet} for this reason. 

Dispute no. 16 between C\&W and PSINet was because C\&W dropped its peering agreement with PSINet and due to this C\&W users were not able to access IP addresses on PSINet network \cite{aol-msn}. One more dispute that we have added to this category is no. 18 between BBN/GTE and MCI/Worldcom, the reason of adding this dispute in this category is that this dispute lead to an administrative enquiry of peering agreements of major providers and this in turn lead to the spinoff of Worldcom's InternetMCI division to Cable \& Wireless \cite{nanog}. 

\subsubsection{Unknown Reasons}
There are some disputes in the table for whom we were not able to find any reason like, dispute no. 12 between Cogent and Limelight in Sept'07 \cite{cogent-limelight}, no. 14 between AOL and MSN in Sept'03, the result of this dispute was that AOL users were not able access MSN \cite{aol-msn}. The other disputes collected in this category are Chunghwa Telecom - Taiwan Fixed Network and Digex Inc - AGIS. The battle between Digex and AGIS was over in a period of one week but during this time AGIS customers were not able to access websites that were on the Digex network \cite{cukier1998peering}. We included dispute no. 21 in this category because, although we know the reason for the dispute,i.e., due to publicizing of peering policy but still we don't know the motive behind this. Hence, we put this dispute here.

One interesting dispute for which we couldn't find any reason is the de-peering between Cogent and China Telecom in Mar'12, although the interesting point about this dispute is as mentioned in \cite{renesys} is that this de-peering increased Sprint's revenue because China Telecom is Sprint's customer and decreased Cogent's revenue because it lost multi-homed customer traffic. So, what might have been the reason for this de-peering? We don't know.

A very popular dispute in the table is of between Cogent and Hurricane Electric (HE), this was an IPv6 dispute where Cogent de-peered HE. Finally, HE managed to sort things out by asking Cogent to establish the peering link again \cite{cogent-he}.

Now that we have explained all the disputes collected in detail, in the next section we'll explain the inferences we make from these disputes.

\section{Results}
\label{results}

\subsection{Based on Timeline}
\begin{figure}[!h]
\centering
\includegraphics[width=2.5in]{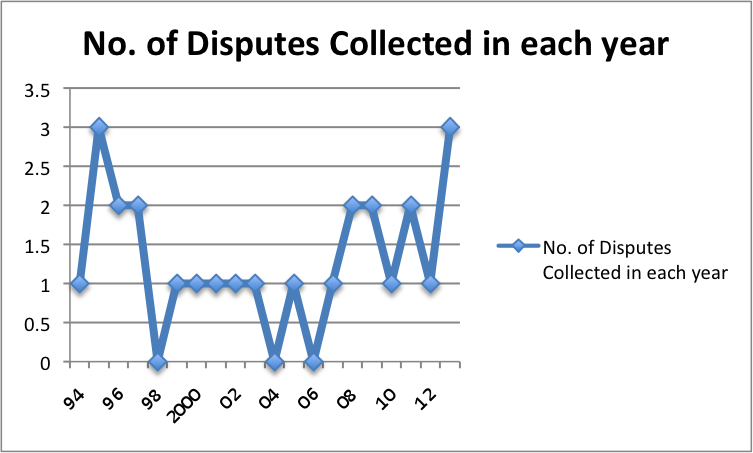}
 \caption{No. of Peering Disputes collected in each year}
\label{timeline}
\end{figure}

Fig. \ref{timeline} represents the disputes collected year wise.  
Looking at Fig. \ref{timeline} it can be noticed there are more number of disputes until 1998, at the rate of at least once per year. There are lesser number of collected disputes during 1999 to 2006 and then again it rises. We have no data of peering disputes in '98 , '04 and '06. This doesn't mean that no peering disputes have taken place in these durations. Typically, such disputes are closed door, proprietary agreements, the details of peering arrangements are handled as trade secrets, and no party benefits from publicly airing grievances. 


If we nibble a little bit more on this then we can see that the increase in number of players can be one of the reasons for the increase in frequency of peering disputes. The earlier peering disputes like, that of Sprint, BBN and AGIS with other ISPs \cite{nanog} were mainly due to technical constraints or issues at the CIX Router and the disputes of the modern time like, that of Cogent with Sprint \cite{sprint-cogent}, Telia \cite{brown2008peering} and Level3 \cite{level3-cogent} \cite{level3-cogent1} are due the imbalanced traffic ratio. This shows that the spotlight has shifted from technological constraints to economics. Technology is no longer a constraint as it used to be and now the major fight is for the share of the revenue that others are generating from the concerned peering link.

\subsection{Geographical Location}
Based on the geographical location we have classified peering disputes data in 5 regions. These regions are based on Regional Internet Registries, i.e. Asia Pacific (APNIC), Europe and Middle East (RIPE NCC), Africa (AfriNIC), Latin America (LACNIC) and North America (ARIN). 
\begin{figure}[!h]
\centering
\includegraphics[width=2.5in]{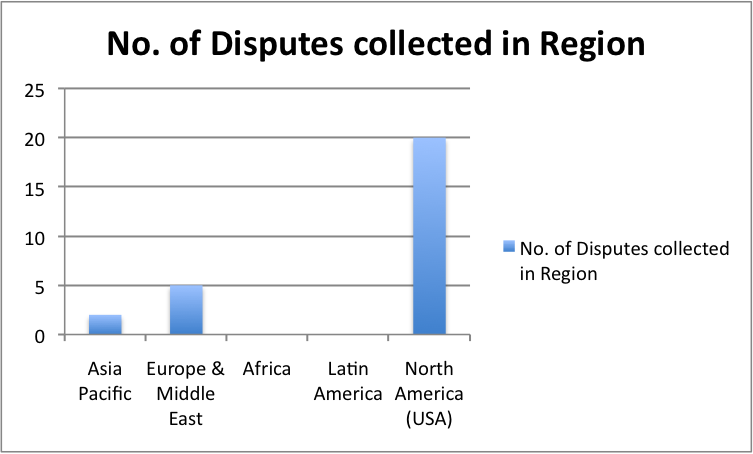}
 \caption{No. of Peering Disputes based on Geographical Location}
\label{fig_sim}
\end{figure}

From Fig. \ref{fig_sim}, it is safe to say that most of the peering disputes are concentrated in the region of North America or to say in a more precise manner, the United States. One can make many inferences out of this like, Does this mean that there are less number of disputes in other regions? or Does this mean that Internet Activity is less in other regions? We have no answer to such questions right now. To be on the safer side, we assume that they are missing from our data.

The reason according to us for this polarity is two fold, one is that Internet originated in United States, i.e. the core of the Internet was formed there and others just joined an already existing network, the presence of majority of Tier-1 Ases (ASes that are on the top of AS hierarchy and do not pay to anyone for transit) in the region makes it essential for others to peer with these existing Tier-1 ASes in US itself, so the possibility of finding peering disputes is more in this part of the globe than in other regions. The other reason is that operators there can afford to get into battles without the interference from the government and the awareness of people for the news of this kind. In \cite{cogent-battle} it is explained that how Cogent is gearing up for another peering battle. The dispute has not even started and already its a news, this shows the eagerness of people to follow such kind of news in the region.

Peering takes place among the ASes of same Tier and if there is a dispute between two Tier -1 ASes then this means that Internet connectivity is disrupted in some part of the world. In \cite{brown2009internet}, it is explained that peering battle number 1 between Cogent and Sprint lead to the breakdown of connectivity for many of their customers (214 ASes were single-homed behind Sprint and 289 were single-homed behind Cogent), and when such large number of ASes loose their connectivity it is bound to become news but why only in the North American region? This is because of the simple fact that most of the Tier-1 ASes are based in this region so if there is a dispute of this scale then it is a burning issue in the region.

Please take a note that a dispute no. 11 between Cogent and Telia has been included in two labels of the bar graph, Europe and North America for the reason that this dispute had its effects in both the regions \cite{brown2008peering}, on the contrary dispute no. 3 and 5 between Cogent and France Telecom (now Orange S.A.) have been included only in European part because the reason for these disputes was the customer base of France Telecom \cite{cogent-ft} \cite{cogent-orange}.

It may be noted that in the bar graph there are no disputes mentioned in the region of Africa and Latin America, this should not be taken as fact that no disputes took place in these regions. We could not find them maybe because they were not publicized or they were in the regional languages. We put in our best efforts to find as many disputes by searching in the regional languages but for these two regions we were not able to find any dispute.

\subsection{Players Involved}
\begin{figure}[!h]
\centering
\includegraphics[width=2.5in]{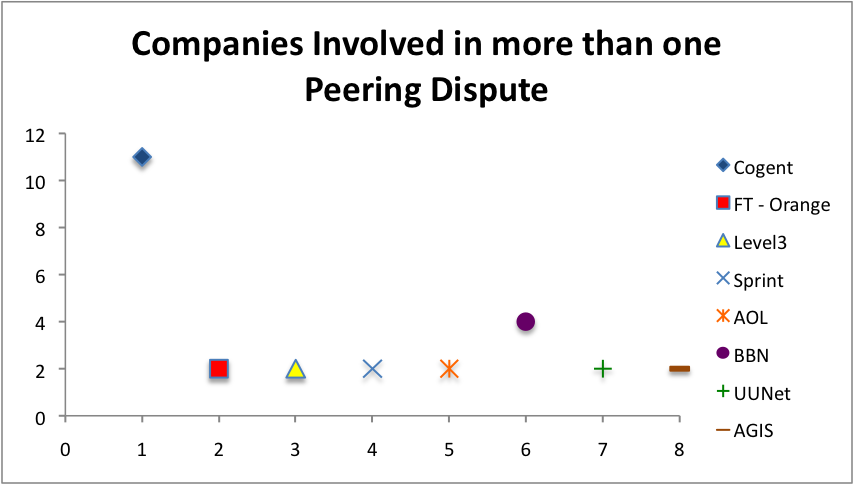}
 \caption{Players involved in more than one Peering Dispute}
\label{fig_players}
\end{figure}

In the TABLE I there are some players that are involved in more then one dispute. In Fig. \ref{fig_players}, we have classified these players based on the number of disputes they are involved in. Please note that we have considered only those companies that are involved in more then one dispute based on the data collected. The points in the graph denote the number of disputes each of these players are involved in.

Cogent that is involved in maximum of the peering disputes collected by us carries a reputation of getting into peering battles more often then others. So whenever there is a peering controversy the first question that is being asked very frequently is, "Is Cogent involved?" As mentioned in \cite{cogent-limelight} Cogent is best known for two things, cheap bandwidth and peering controversies. But this certainly doesn't make Cogent a bad player in the market, one view point of this can be that if you are providing a service at cheaper rates then others then you are bound to get in fights, it's a competitive market out there.

There are players in Fig. \ref{fig_players} that were actually involved in a dispute with each other, the two disputes in which FT-Orange was involved in are with Cogent, so these two disputes have been counted twice.

It is evident that peering disputes are increasing gradually \cite{cs.columbia} more and more players are getting involved because boundaries are changing, there was a time when there were clear distinctions between access ISPs, transit or backbone ISPs, content providers and content delivery networks but over the course of time these distinctions are going extinct, now many companies are in different fields and this has made the market very competitive and this makes these disputes even stronger. The same argument is supported by the statement of Cogent CEO, Dave Schaeffer in \cite{cogent-battle} where he explains that why his company is involved in so many of peering disputes, unlike other companies that now exist in different fields like, voice, wavelength services, private line etc.  So, if you are in field where every player is so diversified, disputes are inevitable. 

\subsection{Impact of depeering}
\label{sec_depeer}
Once a depeering occurs, BGP starts announcing the new reachability information. Thus, routing is in transient state until all the ASes converge to the newly formed topology of the Internet. This finally leads to frequent BGP announcements, routing loops during the convergence of BGP, etc. Note that disputes are a bigger problem up in the hierarchy of ASes. For example, a dispute at tier-1 is going to be visible to many small ISPs that take their services to connect to the Internet. This also forces changes in routing paths and the packets may take really long paths or may not even reach to the destination even if the paths exist. It is inherent to how BGP works.

\section{Conclusion and Future Work} 
\label{conclusion}
Our work is a first step in the direction of systematically analyzing the disputes and classifying them. We collected over 26 disputes. We classified the disputes on grounds like geographical considerations and time line to see how the disputes have varied over different regions and time respectively. We had also came across a dispute that is said to be the first commercial peering dispute, between UUnet and College Park NSS in 1989-90, even before the existence of CIX \cite{1989}. We expect that there are more disputes like this one, which are missing from our data. In the future, we would like to collect more disputes to make an exhaustive list. Depeering has lot of aftermath once it takes place. It also shapes the behavior of ISPs. The detailed study of post peering disputes is left as future work.

\section*{Acknowledgment}
The authors would like to thank everyone who responded to our queries on the mailing lists, particularly NANOG.

\bibliographystyle{IEEEtran}
\bibliography{peering}

\end{document}